\documentclass{iau}

\usepackage{amsmath}
\usepackage{natbib,aas_jrnl_names}
\usepackage{graphicx}
\usepackage{multirow}
\newcommand{\lr}{$\mathrm{L_R}$}

\newcommand{\lx}{$\mathrm{L_X}$}
\newcommand{\vcme}{$\mathrm{V_{CME}}$}
\newcommand{\rec}{$\mathrm{\phi_{rec}}$}

\newcommand{\Pfl}{$\mathrm{\sqrt{L_X\phi_{rec}}}$}
\newcommand{\Pcme}{$\mathrm{\sqrt{L_RV^2_{CME}}}$}
\newcommand{\Pf}{$\mathrm{P_{flare}}$}
\newcommand{\Pc}{$\mathrm{P_{CME}}$}
\newcommand{\cor}[2]{C$_\mathrm{#1,#2}$}
\newcommand{\crc}{\cor{R}{C}}

\begin{document}

\aopheadtitle{Proceedings of the IAU Symposium}
\editors{N. Gopalswamy,  O. Malandraki, A. Vidotto \&  W. Manchester, eds.}

\lefttitle{Mohan, A., et al.}
\righttitle{A multi-mission DH type-II catalog: Initial results}

\jnlPage{1}{7}
\jnlDoiYr{2024}
\doival{10.1017/xxxxx}
\volno{388}
\pubYr{2024}
\journaltitle{Solar and Stellar Coronal Mass Ejections}

\title{A catalog of multi-vantage point observations of type-II bursts: Statistics and correlations}
\author{Atul Mohan$^{1,2}$, Nat Gopalswamy$^1$, Hemapriya Raju$^3$ and Sachiko Akiyama$^{1,2}$}
\affiliation{{}$^1$NASA Goddard Space Flight Center, 8800 Greenbelt Road Greenbelt, MD, 20771, USA}
\affiliation{{}$^2$The Catholic University of America, 620 Michigan Avenue, N.E. Washington, DC 20064, USA}
\affiliation{{}$^3$Indian Institute of Technology, Indore, Simrol, Indore, 453552, India}

\begin{abstract}
Coronal mass ejection (CME) often produces a soft X-ray (SXR) flare associated with the low-coronal reconnection and a type-II radio burst associated with an interplanetary (IP) CME-shock. SXR flares and type-II
bursts outshine the background emission, making them sun-as-a-star observables. Though there exist SXR flare catalogs covering decades of observations, they do not provide the associated type-II luminosity. Besides, since radio burst emission could be beamed, the observed flux dynamic spectrum may vary with line of sight. Using long-term calibrated decameter-hectometric
dynamic spectra from the Wind and STEREO spacecraft, we build a catalog of
multi-vantage point observations of type-II bursts. Cross-matching with existing
catalogs we compile the properties of the associated flare, reconnection, and the CME. Cross-correlation analysis was done between various parameters. Two novel metrics of flare and CME power show a strong correlation revealing a link between particle acceleration strengths in the low corona and IP space. 
\end{abstract}

\begin{keywords}
Solar coronal mass ejections, Catalogs, Stellar coronal mass ejections
\end{keywords}

\maketitle

\section{Introduction}
Solar coronal mass ejections (CMEs) are a major driver of energetic space weather phenomena, like solar energetic particle events (SEPs), geomagnetic storms etc.
CMEs are often associated with flares across the electromagnetic spectrum, from radio to gamma-rays~\citep[see,][for an overview]{howard23_CME_review}.
The two major processes during a flare-CME event include the reconnection event that accelerates electrons leading to the SXR flare in the low corona, and the CME shock that accelerates particles across the corona and interplanetary space often leading to {type-II radio bursts~\citep[e.g.][]{wild50_bursttypes,Mclean85_book,Gopal11_PREconf}} in meter - kilometer (m-km) band. \cite{gopal18_coronalRcflx_IPburstconnection} showed that the low coronal flare-reconenction flux, \rec, correlates well with the CME kinetic energy revealing a link between the strengths of the low coronal and interplanetary impacts of the active phenomenon.
However, despite decades of observations, we lack an understanding of the relationships, if any, between the observational proxies to the strengths of the particle acceleration in the low corona (peak SXR luminosity, \lx) and interplanetary space (peak type-II luminosity, \lr), and the physical properties of their drivers (i.e., flare and CME).
\begin{figure}[]
\centering
  \includegraphics[height=0.22\textheight, width=0.9\textwidth]{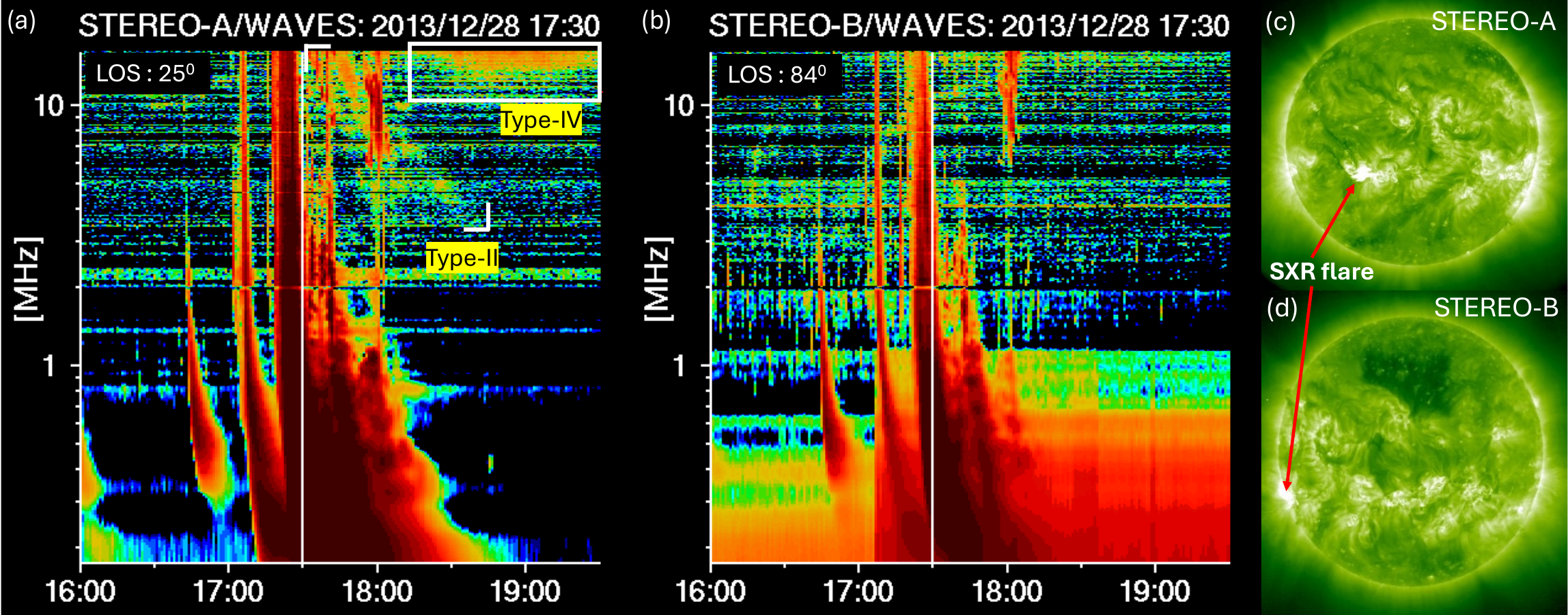}
  \caption{(a-b): DH DS from STEREO-A and B during a CME. The associated active region was at 25$^\circ$ and 84$^\circ$ lines of sight (LOS) in respective spacecraft. The type-II (marked by angular brackets) and type-IV (marked by box) seen in STEREO-A DS is absent in STEREO-B data. (c-d) The 195\AA\ images from both spacecraft show the SXR flare, marked by arrows.}
  \label{fig1:LOSeffect}
\end{figure}

Meanwhile, when it comes to stellar CMEs, unlike the solar case, spatially resolved observations are impossible with current telescopes limiting the capability to directly infer crucial physical parameters, particularly \rec\ and CME speed (\vcme).
Hence, scaling laws linking sun-as-a-star \lx\ and \lr, to relevant spatially resolved physical properties are important to constrain stellar CME models and to understand the solar-stellar flare-CME phenomenon within a common framework.
There exist SXR flare and CME catalogs, that report \lx, \rec\,(in some cases), \vcme\, etc. for decades of solar-CME observations, but they do not report simultaneous \lr\ for any associated type-II bursts.
Besides, m - km burst emission is inherently beamed as demonstrated for different burst types~\citep[e.g.][]{1980A&A....88..218D,Gopal11_PREconf,Hilaire13_typeIIIstats,Atul24_DHtypeIVcatalog}.
Apart from inherent directivity, the line of sight (LOS) plasma and magnetic field structures can also {impact the observed emission characteristics in the dynamic spectrum (DS)}~\citep[e.g.][]{robinson1994,Arzner1999,Nasrin18_DHtypeIV_occult_streamerCMEshock}.
Figure~\ref{fig1:LOSeffect} shows an event observed by STEREO-A and STEREO-B spacecraft with the flaring active region at 25$^\circ$ and 84$^\circ$ lines of sight (LOS) respectively. {The decameter-hectometric (DH) DS and the X-ray images from both spacecraft are compared. 
STEREO-A spacecraft observed the frequency drifting type-II burst and the long-duration stationary type-IV emission confined above 10\,MHz~\citep[see,][for an overview of the burst types]{Mclean85_book}.}
Clearly, STEREO-B does not observe both type-II and type-IV bursts, though both spacecraft see the SXR flare.
Hence the LOS towards the active latitudes on the star can be crucial in choosing a target star with a high chance of detecting a radio burst.
The first step to performing such detailed studies of the properties of the observed type-II bursts is to build a comprehensive catalog of multi-vantage point observations of type-II bursts listing their spectro-temporal properties and \lr, and the properties of the associated flares and CMEs.
We present such a catalog and the initial results on the correlations between various paramters.


\section{Data and event catalog}
The calibrated DH band dynamic spectra recorded by Plasma Wave Investigation (WAVES) instruments onboard Wind, STEREO-A and STEREO-B spacecraft since Nov 2006 form a unique uniformly calibrated multi-vantage point dataset ideal for the goals of the study. This long-term database spans two solar cycles and the rising phase of the current cycle.
The other aspects that make DH bursts interesting are the following. 
DH type-IIs are associated with interplanetary shocks driven by energetic CMEs.
In the stellar CME studies, despite hours and days-long searches on stars with very high flaring rates in the meterwave band, not a single type-II burst has been found~\citep[e.g.][]{bastian90_flarestarsRev,Osten08_ADLeoFinebursts,2018ApJ...856...39C,2019ApJ...871..214V}. 
Numerical simulations suggest that due to the strong magnetic fields in the atmospheres of active stars, an eruptive event may not be able to generate shocks in the lower corona but probably at interplanetary heights, favoring DH band over the meterwave band~\citep{odert20_CME_ObsOverview,alvarado22_CME_star}. 
Also, compared to their metric counterparts, the DH type-II bursts are associated with relatively stronger flares and faster CMEs~\citep{Gopal11_PREconf,Miteva17_SEP-radburst_link}. 
The DH type-II bursts are also strongly associated with SEPs and solar gamma-ray events making them more space weather relevant~\citep{Miteva17_SEP-radburst_link,gopal18_SGRE-typeIIlink}.

\subsection{The multi-vantage point flare- CME- type-II catalog}
The initial list of DH type-II bursts comes from \cite{Gopal19_DHtypeII_catalog}. 
For each event, we assign a `data quality' metric based on the quality of the data from each spacecraft. The data quality metric is of the form WxAyBz, where x,y,z\,$\in$\,{0,1,2}. W stands for Wind, A for STEREO-A and B for STEREO B, while the integer metric attached conveys the quality of the data stream. When 0 represents data gap, 1 represents a poor quality DS in which robust identification of emission features is difficult and 2 represents good data. 
All good dynamic spectra from different spacecraft are inspected and given an `event quality' metric. 
This metric rates the quality of the event recorded. An event is rated along two axes: shape (S) and intensity (I). A type-II event with a well-discernible shape in the DS is given a rating of 3, while 1 represents an event whose
shape is not very evident. This could be because it extends only within a narrow spectral band or because the event is in a crowded region of the DS where other burst types make the shape determination difficult. Similarly along the intensity axis, 3 represents a bright event while 1 denotes an event with a relatively low intensity in the region of the DS where it is detected.
All dynamic spectra with a metric SxIy, where x$>$1 and y$>$1 are generally reliable.
Of all the good dynamic spectra recorded for an event across spacecraft, the DS with the highest event quality metric identifies the observation line of sight that provided the best view of the event. 
The DS with the best event quality is used to estimate the \lr.
The \lr\ was computed as the peak burst flux in the 3 - 7\,MHz range. 
This frequency range was chosen because the type-III and type-II bursts are the major burst types in this range, and are relatively easily separable from one another due to their dynamic spectral morphology.
Most long-duration broadband emission features that last for several hours to days, like the type-IV bursts are mostly confined above 7\,MHz~\citep{Hillaris16_typeIV,Atul24_DHtypeIVcatalog}. Hence in the higher frequency ranges often the type-II emission is mixed with type-IV bursts.
\begin{figure}[]
  \includegraphics[height=0.15\textheight, width=\textwidth]{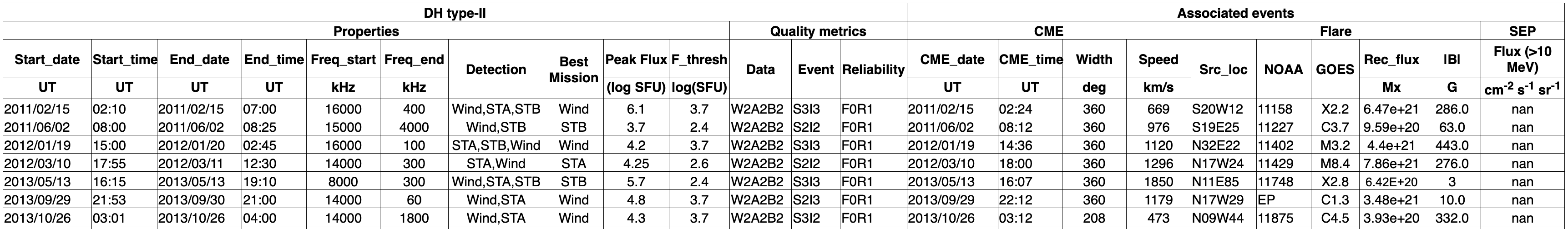}
  \caption{A portion of List 2, that presents DH type-II events with reliable \lr,\lx,\rec\ and \vcme\ estimates. Refer text for details. All event lists and their description can be found at \href{https://cdaw.gsfc.nasa.gov/CME\_list/radio/multimission\_type2/}{https://cdaw.gsfc.nasa.gov/CME\_list/radio/multimission\_type2/}.}
  \label{fig1:eventlist}
\end{figure}

The aforementioned methodology led to a catalog of 163 DH type-II bursts with \lr\ estimates (hereafter `full catalog').
As discussed earlier, since the DS can have multiple co-temporal burst emission features occurring around similar frequency ranges we bestow a `reliability' metric on the \lr\ estimate.
Reliability metric rates \lr\ estimate along two axes: Field crowdedness (F) and relative flux (R). F quantifies how crowded the frequency-time field of the DS is where the type-II is observed. If there are other co-temporal emission features in the 3 – 7 MHz band intervening with the type-II emission, then F is given a high value. A value of 2 is given when the crowdedness is so high that the \lr\ is unreliable, while F=1 implies that \lr\ is reliable but the possibility of contamination from other emission features is non-negligible. 0 denotes the best case where the type-II is not close to any other radio emission feature and hence \lr\ is robust.
Meanwhile, the relative flux (R) metric rates the flux level of the burst in the 3 – 7 MHz range in contrast to the typical burst flux outside of this chosen band. A value of 1 denotes that the 3 - 7\, MHz flux is a good representative of the typical overall DH type-II burst flux level. A value of 0 would mean otherwise. So the F0R1 events have a very reliable \lr, which is also representative of the overall burst flux level.

For each event in the full catalog, \lx\ and \vcme\ were obtained by cross-matching the DH type-II event list with the catalogs maintained by the Coordinated Data Analysis Workshops (CDAW) group\footnote{\href{https://cdaw.gsfc.nasa.gov/}{https://cdaw.gsfc.nasa.gov/}}. Events with good \lx\ and \vcme\ estimates were identified. Among those, the ones with an event quality SxIy, with x,y$\in${2,3} and reliability (F0R1) were chosen and \rec\ was estimated using the post-eruption arcade method~\citep{Gopal17_FREDtech}. 
Finally, a list of 25 `flare- CME- type-II' events (List 2) with reliable \lx,\lr,\vcme, and \rec\ estimates was produced that was used for further analysis. Figure~\ref{fig1:eventlist} shows a portion of List 2. Peak\_Flux provides radio flux density in 3 - 7\,MHz band and F\_thresh the mean background radio flux level. The time and frequency extents of the burst and the various quality metrics are also provided. The `CME' section reports the maximum angular width and the speed of the interplanetary CME along with the time window of the event. Under the `Flare' section, the location of the flaring active region as seen from the Earth (Src\_loc), standard active region ID (NOAA), the peak SXR flux in the GOES scale\footnote{\href{https://www.swpc.noaa.gov/products/goes-x-ray-flux}{https://www.swpc.noaa.gov/products/goes-x-ray-flux}}, \rec\ in Mx unit and absolute magnetic filed strength at the flare region are provided.  
The link to the full catalog and event lists is in the caption of Fig.~\ref{fig1:eventlist}.

\section{Results and discussion}
\begin{figure}[t]
  \centerline{\vbox to 6pc{\hbox to 10pc{}}}
  \includegraphics[scale=.4]{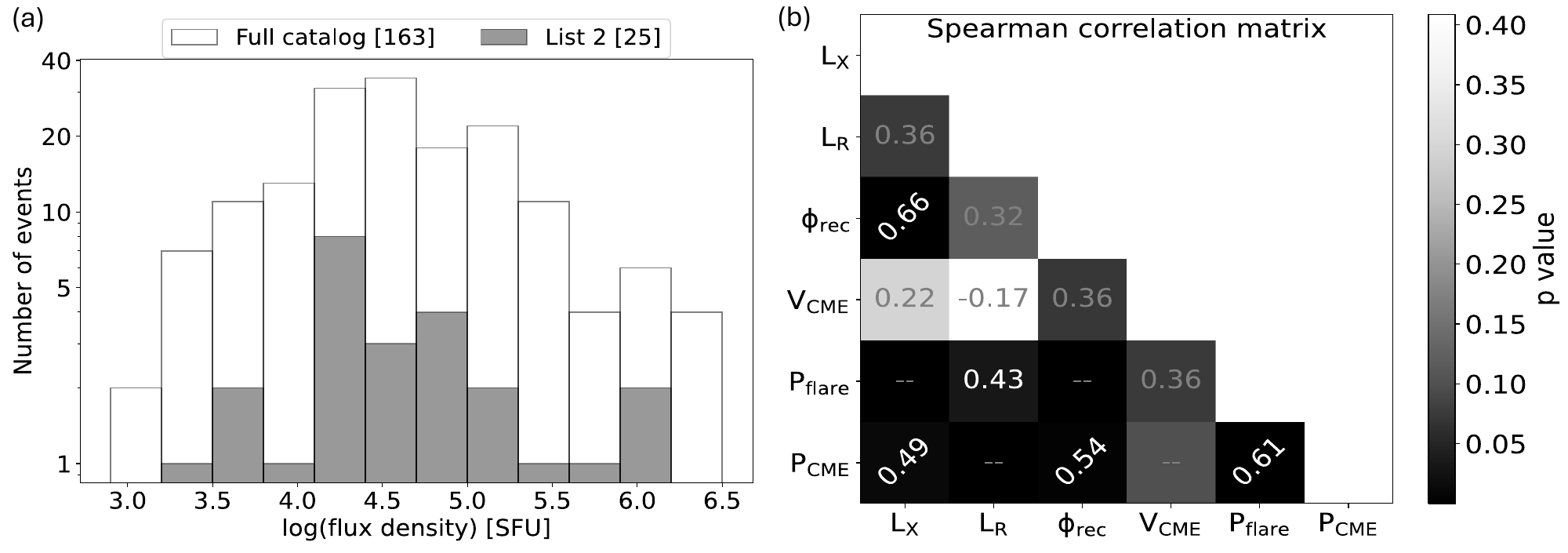}
  \caption{Statistical properties of the flare- CME- type-II events. (a): Flux density histogram. (b): Spearman correlation matrix for the various properties of the List 2 events. Numbers in each cell denote the correlation coefficient \crc\ corresponding to the row (R) and column (C) parameters. Rotated text highlights strong correlations, white normal text denotes reliable weak \crc, and gray denotes the insignificant ones. Whenever R parameter is a function of C, \crc\ is not mentioned.}
  \label{fig1:stats}
\end{figure}
Figure~\ref{fig1:stats}a shows the flux density distribution of the events in the full catalog and the sub-lists. A mean flux density of $\sim$3.5$\times 10^4$\,SFU is found for all lists. Besides \lx,\lr,\vcme\ and \rec\ we define two new metrics of flare and CME power combining these quantities.
The `flare power', \Pf = \Pfl, combines the strength in the magnetic reconnection with that in flare emissivity. Meanwhile the `CME power', \Pc = \Pcme, estimates the combined power in the CME kinetic energy that drives the shock and the associated particle acceleration. Figure~\ref{fig1:stats}b shows the result from the Spearman correlation analysis performed between parameter tuples drawn from the set comprising of \lr, \lx, \rec, \vcme, \Pf\ and \Pc.
Each cell in the matrix plot shows the correlation coefficient (\crc) between the row (R) and column (C) parameter. The cell color represents the p-value. The significant strong \crc\ are highlighted by slanting text (\crc $\gtrsim$ 0.49 \& p$<0.015$). Straight white text represent reliable weak correlations (\crc $<$ 0.49 \& p$<$0.05) and those in gray are insignificant (p$>$0.05; \crc$<$ 0.49). 
The correlation coefficients between tuples that are related by definition are masked and will not be considered in the analysis. 
After the well-known strongly correlated tuple (\rec,\lx)~\citep{Kazachenko17_Ribbondb,Sindhuja20_Recflx_xflr}, (\Pc,\Pf) is the strongly correlated pair with C$_\mathrm{P_{CME}, P_{flare}}$=0.61.
The other strong correlations (green text) include (\Pc,\lx) and (\Pc,\rec).
Of all the strongly correlated parameter tuples where one parameter is fully flare-related and the other fully CME-related, (\Pc,\Pf) shows the highest \crc\ value.
Note that the individual terms that makeup \Pf\ and \Pc\ are not significantly correlated with each other, making the \Pf\ - \Pc\ correlation robust.
This correlation highlights the connection between the low-coronal and interplanetary particle acceleration events driven by the flare and the CME.
\section{Conclusion}
We present a catalog of decameter-hectometric (DH) type-II bursts with multi-vantage point observations from Wind and STEREO spacecraft. The data covers the interval from Nov 2006 to July 2023. Peak type-II flux in 3-7\,MHz was found for all events along with the time-frequency extent of the burst.
The full catalog has 163 events.
For each type-II event, the catalog specifies the spacecraft that recorded the best view of the burst in terms of the time-frequency extent and morphology in the dynamic spectrum (DS). Also, the flare source location as seen from the Earth is mentioned. The DS of the best-viewed burst is rated along shape and intensity axes to help identify the events that demonstrated a well-characterizable shape with sufficient intensity. Along with the information on the source location and the relative position of the STEREO spacecraft, available from online resources\footnote{\href{https://stereo-ssc.nascom.nasa.gov/cgi-bin/make\_where\_gif}{https://stereo-ssc.nascom.nasa.gov/cgi-bin/make\_where\_gif}} the catalog can be used to explore the line of sight dependency in the detection of radio bursts.

Cross-matching with the event catalogs compiled by the NASA-CDAW group, the properties of the associated soft X-ray (SXR) flare and interplanetary (IP) CME are compiled.
The flare reconnection flux (\rec), peak SXR luminosity (\lx) and CME speed (\vcme) were either computed or compiled from existing catalogs, for all possible events.
25 events in the full catalog (List 2) had reliable estimates for \lx,\lr,\rec\ and \vcme.
Using List 2 events, various correlations between flare, CME and type-II parameters were explored, to understand the interconnections, if any, between the particle acceleration processes in the low corona and IP space driven by the flare and the CME respectively. 
The newly defined flare and CME power terms, \Pf\,(\Pfl) and \Pc\,(\Pcme) respectively, demonstrate the highest correlation coefficient among all possible parameter tuples considered with one parameter purely flare-related and the other purely CME-related.
The \rec\ and \vcme\ are also correlated with \Pc.

\bibliography{paper_V3}
\bibliographystyle{iau}
\end{document}